\documentclass{iauc} 
\usepackage{graphicx}

\def\Lya{Ly$\alpha$~} 
\def\OM{$\Omega_{\rm m}$~}
 
\def\OB{$\Omega_{\rm b}$~}
\def\LCDM{$\Lambda$CDM~}
\def\gam{$\Gamma_{\rm HI}$~}
\def\OBh{$\Omega_{\rm b}h^{2}$~}
\def\gamn{$\Gamma_{\rm -12}$~}
\def\sig{$\sigma_{\rm 8}$~}

\def\teff{$\tau_{\rm eff}$~}

\def\etal{$\it et$ $\it al.$~}

\pubyear{2005} 
\volume{199}
\pagerange{1--} 
\jname{Probing Galaxies through Quasar Absorption Lines}
\editors{P.R. Williams, C. Shu, and B. M\'{e}nard, eds.}

\title[Constraints on the metagalactic hydrogen ionization rate] 
{Constraints on the metagalactic hydrogen ionization rate from the Lyman-$\alpha$ forest opacity}

\author[J.S. Bolton, M.G. Haehnelt, M. Viel \and V. Springel]   
{James S. Bolton$^1$, Martin G. Haehnelt$^1$, Matteo Viel$^1$ \break \and Volker Springel$^2$}

\affiliation{$^1$Institute of Astronomy, University of Cambridge, Madingely Road, Cambridge, CB3 0HA, UK \break \\ [\affilskip]
$^2$Max-Planck-Institut f\"{u}r Astrophysik, Karl-Schwarzschild-Str. 1, Garching bei M\"{u}nchen, Germany \break}

\begin{document}

\maketitle

\begin{abstract}
  Understanding the sources responsible for reionizing the Universe is
  a key goal of observational cosmology.  A discrepancy has existed
  between the metagalactic hydrogen ionization rate, $\Gamma_{\rm
    HI}$, predicted by early hydrodynamical simulations of the
  Lyman-$\alpha$ forest if scaled to appropriate assumptions for the
  IGM temperature, when compared to values predicted from the
  proximity effect.  We present new estimates for $\Gamma_{\rm HI}$ in
  the redshift range $2<z<4$ based on hydrodynamical simulations of
  the Lyman-$\alpha$ forest opacity.  Within the current concordance
  cosmology, and assuming updated QSO emissivity rates, a substantial
  contribution to the UV background from young star-forming galaxies
  appears to be required over the entire redshift range.  Our results
  are consistent with lower-end estimates from the proximity effect.
  It is also found that the errors on the ionization rate are
  dominated by uncertainties in the thermal state of the intergalactic
  medium and the r.m.s fluctuation amplitude at the Jeans scale.

\end{abstract}

\firstsection

\section{Introduction}

The series of low column density ($10^{12.5} < N_{\rm HI} < 10^{14.5}$
$\rm cm^{-2}$) \Lya absorption features seen blue-ward of the \Lya
emission line in high redshift quasi-stellar object (QSO) spectra
trace the distribution of neutral hydrogen in the intergalactic medium
(IGM).  This forest of absorption lines provides a unique, unbiased
probe of the thermal and ionization history of the IGM along the QSO
line-of-sight.  This has motivated many authors to use hydrodynamical
simulations of structure formation, calibrated to reproduce the
observed properties of the forest obtained from high resolution QSO
spectra, to infer the amplitude of the metagalactic UV background (see
the seminal paper by Rauch \etal 1997, and most recently Tytler \etal
2004).  Consequently there are a wide range of estimates for the
hydrogen ionization rate per atom, \gam, inferred from the \Lya forest
opacity using numerical simulations with different assumptions.  This
complicates the comparison with determinations of the ionization rate
from estimates of the integrated emission from observed QSOs and/or
galaxies ({\it e.g.}  Haardt \& Madau 1996, hereafter HM96; Madau,
Haardt \& Rees 1999) and estimates using the proximity effect ({\it
  e.g.}  Scott et al. 2000) which both also have rather large
uncertainties.  Furthermore, Steidel \etal (2001), in a study of Lyman
break galaxies at $z \simeq 3.4$, suggested that the intensity of the
ionizing background may be a factor of a few larger than in the model
of HM96 due to a large contribution from star-forming galaxies.

We discuss new estimates of the metagalactic hydrogen ionization rate
presented in Bolton \etal (2005), inferred from state-of-the-art
hydrodynamical simulations of the \Lya forest.  These estimates are
consistent with a substantial contribution from galaxies to the
amplitude of the UV background at the hydrogen ionization edge, and
are in agreement with other observational estimates of the
dimensionless quantity \gamn$=\Gamma_{\rm HI}/10^{-12} \rm s^{-1}$.

\section{Numerical code and convergence}

We use a suite of $21$ high resolution hydrodynamical simulations run
using a new version of GADGET (Springel, Yoshida \& White, 2001) to
investigate in detail the dependence of \gam on physical and numerical
parameters (for a detailed description of the simulations see Bolton
\etal 2005).  The cosmological parameters for our adopted fiducial
model are:

\[ (\Omega_{\rm m},\Omega_{\rm b}h^{2},h,\sigma_{8}) = (0.26\pm0.04,0.024\pm0.001,0.72\pm0.04,0.85\pm0.05), \]

\noindent consistent with the results of Spergel et
al. (2003), assuming a flat, vacuum energy dominated cosmological
model.  We assume $n=0.95$ and a helium mass fraction of $Y=0.24$.  In
addition, we adopt fiducial values for the thermal state of the gas at
mean cosmic density; $T_{0}=[11200,17800,12500]\pm5000$ $\rm K$ at
$z=[2,3,4]$ and $\gamma=1.3\pm0.3$ for the index of the gas effective
equation of state (Hui \& Gnedin 1997), based on the results of Schaye
\etal (2000).  We linearly rescale the opacity of the artificial \Lya
forest spectra we construct to match the central values for the \Lya
forest effective optical depth taken from the fitting formula of
Schaye \etal (2003): $ \tau_{\rm eff} =
[0.130\pm0.021,0.362\pm0.036,0.805\pm0.070]$.  We treat the amplitude
of the UV background as a free parameter, so rescaling the neutral
hydrogen opacity corresponds to linearly rescaling the hydrogen
ionization rate by the same amount.

We initially run $7$ simulations of our fiducial model with differing
box size and resolution to assess the degree of numerical convergence.
We find simulations with less than $400^{3}$ gas particles within a
$30h^{-1}$ comoving Mpc box do not achieve adequate numerical
convergence.  We take our fiducial model to have $200^{3}$ gas
particles within a $15h^{-1}$ comoving Mpc box; we estimate the value
of \gamn we infer will be systematically high by around $8$ per cent
due to the combined error from box size and resolution.

\begin{figure}
  \includegraphics[width=1.0\textwidth]{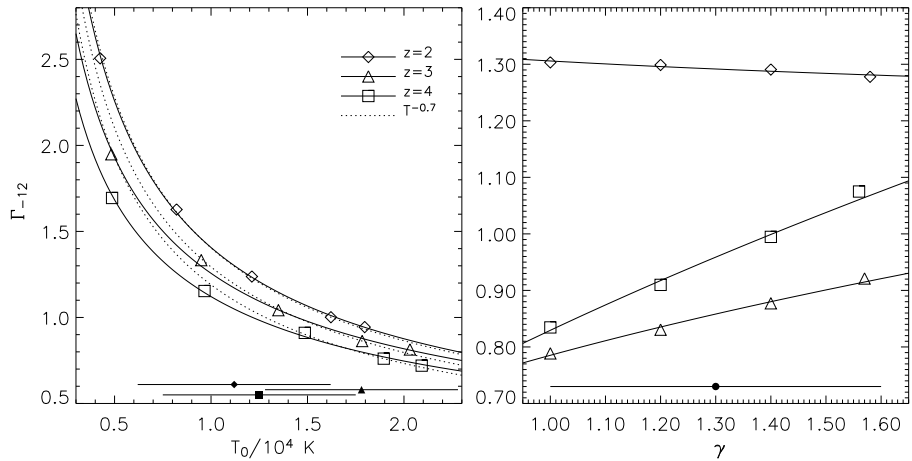}
  \caption{{\it Left:} The dependence of $\Gamma_{-12}$ on the gas temperature
    at mean density $T_{0}$ at the three different redshifts indicated
    on the plot.  Solid curves show a least squares fit and the dotted
    curves shows the $T^{-0.7}$ scaling due to the temperature
    dependence of the recombination coefficient.  The filled symbols
    show the fiducial temperatures and their uncertainties.  {\it
      Right:} The dependence of $\Gamma_{-12}$ on the index of the
    temperature density relation $\gamma$.  The filled circle shows
    our fiducial value of $\gamma$ and its uncertainty.}
 \label{fig:fig2}
\end{figure}

\section{Scaling relations for $\Gamma_{-12}$ from simulated absorption spectra}

When inferring \gamn from hydrodynamical simulations and subsequently
rescaling with different parameters, it is generally assumed that
($e.g$ Rauch \etal 1997):

\begin{equation} \Gamma_{-12} \propto  \Omega_{\rm b}^{2}h^{3}T^{-0.7}\Omega_{\rm m}^{-0.5}.  \label{eq:gscale} \end{equation}

\begin{figure} 
  \includegraphics[width=1.0\textwidth]{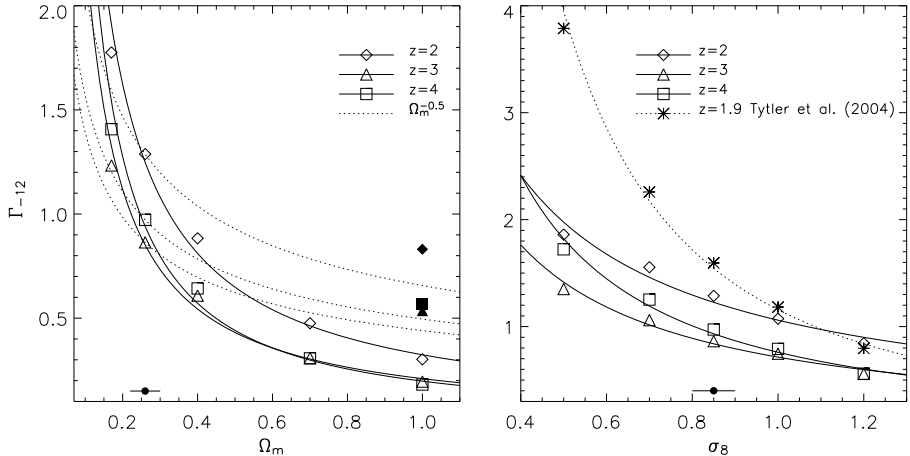}
  \caption{ 
    {\it Left:} The dependence of the estimated $\Gamma_{-12}$ on
    $\Omega_{\rm m}$ for the three different redshifts indicated on
    the plot.  The solid curves are a least-square fit.  The dotted
    lines show the $\Omega_{\rm m}^{-0.5}$ scaling. Filled points are
    obtained for a model with $\Omega_{\rm m}=1.0$ and the the same
    r.m.s.  fluctuation amplitude at a scale of $30$ $\rm kms^{-1}$ as
    our fiducial model.  The filled circle shows our fiducial value of
    $\Omega_{\rm m}=0.26$ and its uncertainty.  {\it Right:} The
    dependence of $\Gamma_{-12}$ on $\sigma_{8}$.  The dotted curve
    shows the result of Tytler et al.  at $z=1.9$, assuming $\langle F
    \rangle_{\rm obs} = 0.882 \pm 0.01$.  The filled circle shows our
    fiducial value of $\sigma_{8}$ and its uncertainty}
\label{fig:fig3}
\end{figure}

It is implicit in this relation that the density and velocity
distribution, along with the effective equation of state of the \Lya
absorbers remain unchanged for different values of $\Omega_{\rm b}$,
$h$, $\Omega_{\rm m}$, $T$ and $\Gamma_{-12}$.  We test the validity
of equation~\ref{eq:gscale} using $14$ simulations in addition to our
fiducial model, by varying the parameters $T$, \OM and \sig in each
run.  We also rescale to different values for the temperature-density
relation index $\gamma$ and the effective optical depth in
post-processing.  We do not discuss the scaling with \OB and $h$,
since our analysis revealed that equation~\ref{eq:gscale} holds
extremely well for these parameters.  Figures~\ref{fig:fig2} and
~\ref{fig:fig3} and Table~\ref{tab:param3} summarise the main results
of this study.  We find that \gamn scales around our fiducial model
as:

\begin{equation} \Gamma_{-12} \propto \Omega_{\rm b}^{2}h^{3}T^{x_{1}(z)}\gamma^{x_{2}(z)}\Omega_{\rm m}^{x_{3}(z)}\sigma_{8}^{x_{4}(z)}\tau_{\rm eff}^{x_{5}(z)},  \label{eq:totscale} \end{equation}

\begin{table} 
\centering
  \caption{
    The redshift dependent indices from our scaling relation of $\Gamma_{-12}$ with several cosmological and astrophysical parameters.}
  \begin{tabular}{cccccc}

    \hline
         & $T$ & $\gamma$ & $\Omega_{\rm m}$ & $\sigma_{\rm 8}$ & $\tau_{\rm eff}$ \\
    \hline
    $z$  & $x_{1}(z)$ & $x_{2}(z)$ & $x_{3}(z)$ & $x_{4}(z)$ & $x_{5}(z)$ \\        
  \hline
     2   & -0.68 & -0.04 & -1.00 & -0.90 & -1.44 \\
     3   & -0.62 &  0.34 & -1.04 & -0.99 & -1.61 \\
     4   & -0.59 &  0.55 & -1.16 & -1.26 & -1.68 \\
  \hline
  
  \label{tab:param3}
\end{tabular}
\end{table}

\noindent
where we tabulate the indices of equation~\ref{eq:totscale} in
Table~\ref{tab:param3}. We stress that this scaling relation is likely
to be somewhat model dependent and should not be applied to models
with parameters very different from our fiducial model without further
checks.  However, it does provide a clear picture of the degeneracies
which exist between \gamn and other parameters within our simulations,
and is independent of assumptions about the gas distribution,
effective equation of state and ionized gas fraction of the low
density IGM.  In particular, the \Lya optical depth at fixed r.m.s.
fluctuation amplitude \sig is more strongly dependent on \OM than
equation~\ref{eq:gscale} suggests due to changes in the gas density
distribution.  Thermal broadening also produces a deviation from the
scaling of equation~\ref{eq:gscale} for temperature, although this
change is much less dramatic.  Consequently, we urge caution when
using equation~\ref{eq:gscale}, especially if comparing models with
differing $\Omega_{\rm m}$.  We also find a strong dependence of \gamn
on the effective optical depth of the \Lya forest which the simulated
spectra are scaled to match.

\section{Results}

Using our scaling relations, we estimate the error on the values of
\gamn from our fiducial simulation using the uncertainties on our
fiducial parameter values taken from the literature.  The error budget
is listed in Table~\ref{tab:errors}.  We find the largest contribution
is from the uncertainty in the gas temperature.  There is also a large
contribution from the uncertainty in the effective optical depth,
especially at $z=2$.  Interestingly, the uncertainty in \OM also gives
a substantial contribution to the total error budget.  This is
primarily due to the sensitivity of \gamn on \OM due to changes in the
r.m.s.  fluctuations of the gas density at the Jeans scale.  Other
uncertainties are less important, in particular $\Omega_{\rm b}h^{2}$,
\sig and $h$.  Using our fiducial model, we find the \Lya effective
optical depth of the IGM at $z=[2,3,4]$ is reproduced by
\gamn$=[1.29\pm^{0.80}_{0.46},0.86\pm^{0.34}_{0.26},0.97\pm^{0.48}_{0.33}]$.
 
\begin{figure}
  \includegraphics[width=1.0\textwidth]{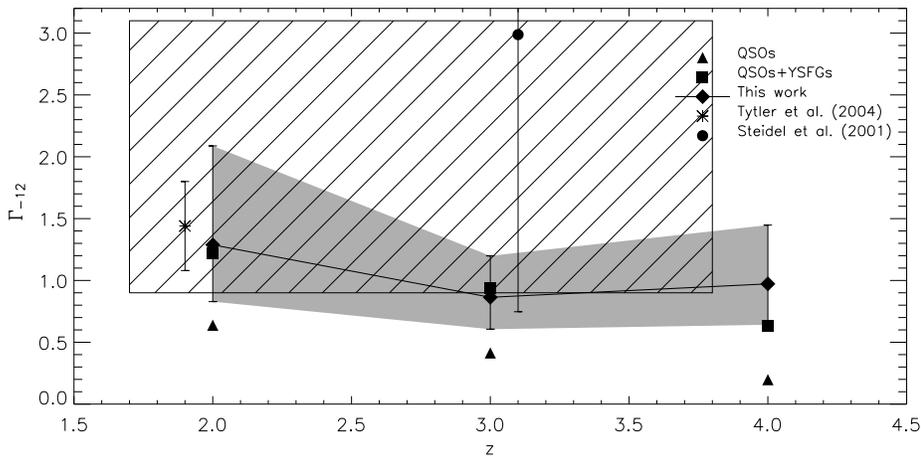}
  \caption{ 
    Comparison of our best estimate for $\Gamma_{-12}$ at $z=[2,3,4]$
    with constraints from observations.  Our data are plotted with
    filled diamonds, and the grey shaded area shows the error bounds.
    The filled squares and triangles show the estimated contribution
    to the metagalactic ionization rate from QSOs+galaxies and QSOs
    alone, based on estimates from the updated model of Madau, Haardt
    \& Rees (1999) including UV photons from re-processing by the IGM.
    The hatched box gives the constraint on \gamn from the proximity
    effect (Scott \etal 2000) and the estimate from Lyman-break
    galaxies assuming a global spectral index of $\alpha=1.8$ is
    plotted with a filled circle (Steidel \etal 2001).  The data point
    has been offset from $z=3$ for clarity, and the upper error limit
    is not shown.  The star shows the best estimate of \gamn at
    $z=1.9$ from Tytler \etal (2004), including their errors from the
    uncertainty in $\tau_{\rm eff}$, \sig and $\Omega_{\rm b}$.}
\label{fig:fig7}
\end{figure}

We compare our results with other observational estimates of \gamn in
fig.~\ref{fig:fig7}.  The filled triangles show the metagalactic
hydrogen ionization rate computed from the updated UV background model
of Madau, Haardt \& Rees (1999), hereafter MHR99 (Francesco Haardt,
private communication), based on the contribution from QSOs and
re-processing by the IGM ({\it e.g.} HM96).  Our data appear to be
inconsistent with the IGM being kept ionized by QSOs at $z\leq4$.  For
comparison, the filled squares give \gamn calculated using the QSO
rate above, plus an additional source in the form of Young Star
Forming Galaxies (YSFGs).  These ionization rates are in good
agreement with our results, suggesting that a substantial contribution
from galaxies appears to be required at all redshifts.

Constraints on \gamn from the proximity effect measurements of Scott
\etal (2000) are shown by the hatched box.  Our results are consistent
with the lower end of these estimates.  The filled circle gives the
\gamn we calculate assuming a spectral index of $\alpha=1.8$ using the
metagalactic ionizing radiation intensity inferred from Lyman-break
galaxies (Steidel \etal 2001).  Finally we also plot the value of
\gamn inferred by Tytler \etal in their study of the \Lya forest at
$z=1.9$.  The error bars include the uncertainties in $\tau_{\rm
  eff}$, \sig and $\Omega_{\rm b}$, based on scaling from their
hydrodynamical simulations.  Our data is in reasonable agreement with
the Tytler et al. result, allowing for differences between the exact
gas temperatures and numerical method.

\begin{table} 
\centering
\caption{
  Percentage error budget for $\Gamma_{-12}$ from estimates of various cosmological and astrophysical parameters, listed approximately in order of importance.  The total error is obtained by adding the individual errors in quadrature.} 

\begin{tabular}{ccccc}
  \hline
    Parameter & $z=2.00$ & $z=3.00$ & $z=4.00$ \\  
    \hline
    $T$           & $^{+50}_{-22}$ & $^{+23}_{-14}$ & $^{+35}_{-18}$   \\
    \OM           & $^{+18}_{-13}$ & $^{+19}_{-14}$ & $^{+21}_{-15}$   \\
    \teff         & $^{+29}_{-19}$ & $^{+18}_{-14}$ & $^{+17}_{-13}$   \\
    Numerical     &  $\pm$10$$ &  $\pm$10$$ &  $\pm$10$$  \\  
    $\gamma$      & $\pm$1$$ & $^{+7}_{-9}$ & $^{+12}_{-13}$   \\
    \OBh          & $^{+9}_{-8}$ & $^{+9}_{-8}$ & $^{+9}_{-8}$   \\
    \sig          & $^{+6}_{-5}$ & $\pm$6$$ & $^{+8}_{-7}$   \\
    $h$           &  $\pm$6$$ &  $\pm$6$$ &  $\pm$6$$ \\ 
    \hline
    Total & $^{+62}_{-36}$ & $^{+39}_{-30}$ & $^{+49}_{-34}$   \\
    \hline  
    \label{tab:errors}
  \end{tabular}  
\end{table}

\section{Conclusions}

In recent years, compelling observational evidence has led to the
acceptance of a standard cosmological model which is flat, has low
matter density and a substantial contribution of vacuum energy to the
total energy density. Within this 'concordance' cosmological model the
current generation of hydrodynamical simulations predict values for
the metagalactic hydrogen ionization rate, required to reproduce the
effective \Lya optical depth of the IGM in the range $z=2-4$, which
are about a factor of four larger than those in an Einstein-de Sitter
model with the same r.m.s. density fluctuation amplitude $\sigma_{8}$.
The ionization rates estimated from the \Lya forest opacity are more
than a factor two larger than estimates from the integrated flux of
optically/UV bright observed QSOs alone.  This discrepancy increases
with increasing redshift.  We confirm the findings of Tytler et al.
(2004) at $z \sim 1.9$ that the estimated ionization rates from
simulations of a \LCDM concordance model are in reasonable agreement
with the estimates of the integrated ionizing flux from observed QSOs
plus a significant contribution from galaxies as in the model of
MHR99.  The estimates of the ionization rate are also in agreement
with the lower end of the range of values from the proximity effect.
This new agreement appears to be in part due to the currently favoured
low value of \OM, which increases the required value of \gam, and more
accurate measurements of the \Lya forest opacity.  We also find that
the uncertainty on the magnitude of \gam inferred from simulations is
greater than previously estimated, with the primary contribution to
this error coming from the temperature of the low density gas in the
IGM.  Better constraints on the thermal history of the IGM are
required.  Additional physics such as radiative transfer, galactic
feedback and metal enrichment may need to be incorporated into
simulations in a more realistic fashion.  However, our estimate of the
error on \gamn most likely accounts for the modest changes expected
from these processes, suggesting we have obtained a consistent
constraint on the metagalactic hydrogen ionization rate.

\begin{acknowledgments}
  We are grateful to Francesco Haardt for making his updated UV
  background model available to us.  The simulations were run using
  the Altix 3700 supercomputer COSMOS at the Department of Applied
  Mathematics and Theoretical Physics in Cambridge.  COSMOS is a
  UK-CCC facility which is supported by HEFCE and PPARC.  We
  acknowledge support from the European Community Research and
  Training Network ``The Physics of the Intergalactic Medium'', NSF
  Grant No. PHY99-07949 and PPARC. JSB and MV also thank the IAU for
  financial assistance.
\end{acknowledgments}

\end{document}